\documentclass{article}%
\usepackage{amsmath}
\usepackage{amsfonts}
\usepackage{amssymb}
\usepackage{graphicx}%
\providecommand{\U}[1]{\protect\rule{.1in}{.1in}}

\begin{document}

\author{Giuseppe Castagnoli\thanks{giuseppe.castagnoli@gmail.com}}
\title{Mechanism of the quantum speed-up}
\maketitle

\begin{abstract}
We explain the mechanism of the quantum speed-up -- quantum algorithms
requiring fewer computation steps than their classical equivalent -- for a
family of algorithms. Bob chooses a function and gives to Alice the black box
that computes it. Alice, without knowing Bob's choice, should find a character
of the function (e. g. its period) by computing its value for different
arguments. There is naturally correlation between Bob's choice and the
solution found by Alice. We show that, in quantum algorithms, this correlation
becomes quantum. This highlights an overlooked measurement problem: sharing
between two measurements the determination of correlated (thus redundant)
measurement outcomes. All is like Alice, by reading the solution at the end of
the algorithm, contributed to the initial choice of Bob, for half of it in
quantum superposition for all the possible ways of taking this half. This
contribution, back evolved to before running the algorithm, where Bob's choice
is located, becomes Alice knowing in advance half of this choice. The quantum
algorithm is the quantum superposition of all the possible ways of taking half
of Bob's choice and, given the advanced knowledge of it, classically computing
the missing half. This yields a speed-up with respect to the classical case
where, initially, Bob's choice is completely unknown to Alice.

\end{abstract}

\section{Foreword}

Our explanation of the speed-up relies on the interplay between the unitary
part of the quantum algorithm and the initial an final measurement operations.
Because of the interdisciplinary character of the work, we spend a few lines
to introduce the language of quantum computation.

An algorithm is the computation that solves a problem. Quantum computation is
the implementation of the algorithm at a fundamental physical level. A quantum
algorithm yields a speed-up when it requires fewer computation steps than its
classical equivalent, sometimes demonstrably fewer than classically possible.

We focus on a family of quantum algorithms that comprises the major speed-ups.
Bob, the problem setter, chooses a function out of a known set of functions
and gives to Alice a black box that computes it. Alice, without knowing Bob's
choice, should find a character of the function by computing its value for
different arguments.

The essential things of a quantum computation process are:

\begin{enumerate}
\item The \textit{register}, which contains a number or a quantum
superposition thereof. The register's initial state is the input of the
computation; e. g.: $\frac{1}{2}\left(  \left\vert 00\right\rangle +\left\vert
01\right\rangle +\left\vert 10\right\rangle +\left\vert 11\right\rangle
\right)  $.

\item The (reversible) computation, a \textit{unitary transformation} $U$ that
sends the input into the output -- the result of the computation or the
solution of the problem. E. g.: $U\frac{1}{2}\left(  \left\vert
00\right\rangle +\left\vert 01\right\rangle +\left\vert 10\right\rangle
+\left\vert 11\right\rangle \right)  =\frac{1}{\sqrt{2}}\left(  \left\vert
00\right\rangle +\left\vert 11\right\rangle \right)  $.

\item The \textit{initial measurement}, required to prepare the register in
the desired initial state. In particular, we are interested in the preparation
of the choice of the problem on the part of Bob.

\item The \textit{final measurement}, performed by Alice in the output state
of the register to read the solution of the problem.
\end{enumerate}

The seminal speed-up was discovered by Deutsch $\left[  1\right]  $ in 1985.
The subsequent speed-ups can be seen as ingenious mathematical extrapolations
of Deutsch's algorithm. We provide an example of speed-up. Given a chest of
four drawers. Bob hides a ball in one of the drawers, Alice should locate the
ball by opening different drawers. Classically, to be sure of locating the
ball, Alice should plan to open three drawers. Quantally, one drawer. This is
the simplest instance of Grover's $\left[  2\right]  $ quantum search algorithm.

In 2001, Grover $\left[  3\right]  $ called for a two-line explanation of the
reason for the speed-up, one that does not enter into the mathematical detail
of each quantum algorithm. It can be said that quantum computer science
prevailingly explored the opposite direction, focusing on the mathematical
gear of quantum algorithms and trying to unify it. In the context of the
present work, it is important to note that such attempts mostly focus on the
unitary transformation part of quantum algorithms. Apropos of this, it should
be noted that many quantum algorithms are unitary transformations starting and
ending with a sharp state of the register. This might have provided the
feeling that the speed-up is essentially deterministic in character.

In 2009, Gross, Flammia, and Eisert $\left[  4\right]  $ claimed that the
\textit{exact reason} for the quantum speed-up had never been explained.

Our explanation of the quantum speed-up, relying on the interplay between the
unitary and the non-unitary part of the quantum algorithm, goes against the
aforesaid trend. It also shows that the apparent determinism of the quantum
speed-up is a sort of "visual illusion".

\section{Extended summary}

Our explanation of the speed-up starts with the obvious observation that there
is correlation between the problem and its solution -- e. g., one to one
correlation between the drawer number initially chosen by Bob and the drawer
number eventually found by Alice. The explanation can be divided in three
steps. The first step (Section 3.1) is showing that, in quantum algorithms,
this correlation becomes quantum.

We divide the register into sub-registers -- for short "registers" as well. In
the four drawer case, we have a two-quantum-bit (qubit) register $B$, under
the control of Bob, and a two-qubit register $A$, under the control of Alice.
Let $\mathbf{b}$ be the number of the drawer with the ball, $\mathbf{a}$\ that
of the drawer opened by Alice. Bob writes his choice of the value of
$\mathbf{b}$, say $\mathbf{b}=00$, in register $B$. Alice writes the number of
the drawer she wants to open in register $A$. Reading the content of a
register amounts to measuring a corresponding observable. We call $\hat{B}$
the content of register $B$, of eigenvalues $\mathbf{b}\in\left\{
00,01,10,11\right\}  $, and $\hat{A}$ the content of register $A$, of
eigenvalues $\mathbf{a}\in\left\{  00,01,10,11\right\}  $. We note that
$\hat{A}$ and $\hat{B}$ commute.

We assume that register $B$ is initially in a maximally mixed state -- so that
the value of $\mathbf{b}$ is completely undetermined. In fact we want to
examine the entire process that leads to the determination of Bob's choice.

In order to prepare register $B$ with the desired value of $\mathbf{b}$, Bob
should measure $\hat{B}$ in the maximally mixed state of register $B$. He
obtains an eigenvalue at random, say $\mathbf{b}=01$. Then he changes the
corresponding eigenstate into the desired one, by applying to $B$ a suitable
permutation of the basis vectors, namely a unitary transformation $U_{B}$. At
this point Alice runs the unitary part of the quantum algorithm, which
eventually writes the solution in register $A$. Alice acquires the solution
$\mathbf{a}=\mathbf{b}=00$ by measuring $\hat{A}$.

A crucial point of our argument is noting that there is quantum correlation
between the outcome of the initial measurement of $\hat{B}$ and that of the
final measurement of $\hat{A}$. In fact, quantum correlation concerns
repetitions of the same quantum experiment. Therefore, from the standpoint of
it, the unitary transformation $U_{B}$ should be considered fixed. The fact
that Bob chooses $U_{B}$ to always obtain the choice $00$, independently of
the outcome of measuring $\hat{B}$, belongs to a different film. Looking only
the latter film originates the aforesaid "visual illusion" -- the apparent
determinism of quantum computation and the speed-up.

Instead, the quantum speed up essentially relies on (non-deterministic)
quantum correlation. As the fixed permutation of a randomly selected value of
$\mathbf{b}$, Bob's choice $\mathbf{b}=00$ should be considered random. Up to
the fixed permutation $U_{B}$, there is quantum correlation between Bob's
choice $\mathbf{b}=00$ and Alice's reading of the solution $\mathbf{a}%
=\mathbf{b}=00$.

This can be best seen by virtually deferring the measurement of $\hat{B}$ at
the end of the unitary part of the quantum algorithm -- see Section 3.1 for
details. The state of the two registers, before the measurement of either
$\hat{B}$\ or $\hat{A}$, is:%
\begin{equation}
\frac{1}{2}\left(  \operatorname{e}^{i\varphi_{0}}\left\vert 00\right\rangle
_{B}\left\vert 00\right\rangle _{A}+\operatorname{e}^{i\varphi_{1}}\left\vert
01\right\rangle _{B}\left\vert 01\right\rangle _{A}+\operatorname{e}%
^{i\varphi_{2}}\left\vert 10\right\rangle _{B}\left\vert 10\right\rangle
_{A}+\operatorname{e}^{i\varphi_{3}}\left\vert 11\right\rangle _{B}\left\vert
11\right\rangle _{A}\right)  . \label{qcorrelation}%
\end{equation}
The $\varphi_{i}$ are independent random phases, each with uniform
distribution in $\left[  0,2\pi\right]  $. We use the random-phase
representation of a density operator to keep the ket vector representation of
the quantum algorithm. The density operator is the average over all
$\varphi_{i}$ of the product of the ket by the bra: $\left\langle \left\vert
\psi\right\rangle \left\langle \psi\right\vert \right\rangle _{\forall
\varphi_{i}}$. The von Neumann entropy of this state is two bits. Measuring
$\hat{B}$\ or $\hat{A}$ projects (\ref{qcorrelation}) on:%
\begin{equation}
\left\vert 00\right\rangle _{B}\left\vert 00\right\rangle _{A}. \label{bob}%
\end{equation}
Back evolving the projection of (\ref{qcorrelation}) on (\ref{bob}),\ by the
inverse of the time forward unitary transformation, restores the projection of
the initial state of register $B$ on $\mathbf{b}=01$.

Let us sum up the situation. We are dealing with two measurements -- Bob's
measurement of $\hat{B}$ and Alice's measurement of $\hat{A}$ -- whose
outcomes are completely (as in the four drawers case) or partly (more in
general) correlated.

It is of course unquestionable that quantum measurement determines an
eigenvalue of the measured observable, but when there are two measurements for
two redundant or partly redundant eigenvalues, what do we have to say?

Interestingly, while quantum correlation has been the source of an enormous
amount of research, the problem of fairly sharing between two measurements the
determination of two completely (or partly) correlated eigenvalues -- thus
completely (or partly) redundant with one another -- has been overlooked.

To analyze this problem, it is useful to introduce the reduced density
operators of registers $B$ and $A$ in state (\ref{qcorrelation}), respectively
$\rho_{B}$ and $\rho_{A}$. The usual way of solving this problem is thinking
that the measurement performed first takes the lion's share. If we assume that
the measurement of $\hat{B}$ is performed first, we ascribe to it: the zeroing
of the entropy of $\rho_{B}$ and the zeroing (or reduction) of the entropy of
$\rho_{A}$. If we assume that the measurement of $\hat{A}$ is performed first,
we ascribe to it: the zeroing of the entropy of $\rho_{A}$ and the zeroing (or
reduction) of the entropy of $\rho_{B}$. We call these two perspectives "the
lion's share perspectives".\ \ \ \ \ \ \ 

The second step of our explanation (Section 3.2) relies on performing this
sharing in a way \ that is not affected by the order of the two measurement,
or the fact that the two measurement are performed simultaneously.

This is justified by the following consideration. Determination means
reduction of entropy, due to the projection -- associated with quantum
measurement -- of a state of higher entropy on one of lower entropy . However,
while quantum measurements are localized in time, the corresponding
projections are not, they can be back evolved along the unitary part of the
quantum algorithm by the inverse of the time-forward unitary transformation.
Therefore, there is no reason to ascribe the lion's share to the measurement
performed first, not to speak of the fact that the two measurements can be simultaneous.

To present ends, it suffices to focus on sharing -- between Alice's and Bob's
measurements -- the projection on Bob's choice. The physical meaning of such a
sharing is of course to be found in the notion of partial measurement of
$\hat{B}$. For example, we can think of measuring $\hat{B}_{0}$, the content
of the left cell of register $B$, in state (\ref{qcorrelation}). A-proiori,
this yields either $b_{0}=0$ or $b_{0}=1$ ($b_{0}$ being the left bit of
$\mathbf{b}$). In present assumptions, the overall measurement of $\hat{B}$
projects on Bob's choice $\mathbf{b}=00$, we are in fact discussing how to
share this projection. This naturally implies the assumption that the
measurement of $\hat{B}_{0}$ yields $b_{0}=0$.

Summing up, we should divide the projection on Bob's choice into \textit{two
projections}. Each projection is associated with a partial measurement of
$\hat{B}$, with outcome post-selected to match with Bob's choice. One
projection should be ascribed to Bob's measurement of $\hat{B}$, the other to
Alice's measurement of $\hat{A}$.

We define our \textit{sharing rule} as follows. To start with, we get rid of
all redundancy by resorting to \textit{Occam's razor}, or law of parsimony. In
Newton's formulation $\left[  5\right]  $, it states \textquotedblleft%
\textit{We are to admit no more causes of natural things than such that are
both true and sufficient to explain their appearances}\textquotedblright. This
law of parsimony requires (i) that the two projections completely determine
Bob's choice without any over-determination, namely without projecting twice
on the same information.

It is reasonable to require that the two projections keep the common qualities
of the two lion's share perspectives. By this we mean that we require that
each projection "properly" reduces the entropies of both $\rho_{B}$ and
$\rho_{A}$. Moreover, from a quantitative standpoint, we require that entropy
reductions are shared in a way that mirrors the symmetry of the measurement
situation. For example, in the four drawer case -- Eq. (\ref{qcorrelation}) --
this implies that the sharing of entropy reductions between Alice's and Bob's
measurements is fifty-fifty. This is condition (ii).

Eventually, we require (iii) that the sharing of the projection on Bob's
choice is done in all the possible ways, compatible with the former
conditions, in quantum superposition.

The above conditions (i) through (iii) are enough to univocally solve the
"sharing problem" in all the quantum algorithms examined in this paper.

The third step of our explanation (Section 3.3) is showing that Alice's
contribution, back evolved to before running the algorithm (by the inverse of
the time-forward unitary transformation), where Bob's choice is positioned,
becomes Alice knowing half of Bob's choice in advance, in all possible ways in
quantum superposition.

Correspondingly, the quantum algorithm turns out to be the quantum
superposition of all the possible ways of taking half of Bob's choice and,
given the advanced knowledge of this half, classically computing the missing
half. This of course yields a speed-up with respect to the classical case
where Bob's choice, initially, is completely unknown to Alice.

This explanation of the quantum speed-up has two main implications:

(I) The quantum speed-up comes from comparing two classical algorithms, with
and without advanced knowledge of half of Bob's choice. This implication is
thus a tool for finding the achievable speed-ups -- a central problem in
quantum computation. Moreover, this problem is brought to an entirely
classical framework, an important simplification.

(II) It shows that the quantum speed-up hosts a special causality loop. In
quantum search, Alice knows in advance 50\% of the bits that specify the
solution, in all possible ways in quantum superposition, and, given the
advanced knowledge of these bits, reaches the solution with fewer computation
steps. Each individual history contains a causality loop; in the four drawer
case, Alice knows in advance that the ball is in one of two drawers, and this
before opening any drawer. Thus, opening any one of these two drawers allows
Alice to locate the ball. This would be impossible if there were only one
isolated history. It becomes possible because, in the superposition of all
histories, all the drawers are opened and there is cross-talk between
histories because of quantum interference.

In the following Sections, first we develop our argument in detail for
Grover's algorithm, then we show that it holds unaltered for the very diverse
quantum algorithms that yield an exponential speed up.

With respect to Ref. $\left[  6\right]  $, we bring to a fundamental physical
level the problem of sharing between Alice's and Bob's measurements the
determination of Bob's choice. This allows us to extend, to Deutsch and
Jozsa's algorithm, Simon's algorithm and the Abelian hidden subgroup
algorithm, the detailed explanation of the mechanism of the speed-up already
developed for Grover's algorithm.

\section{Grover's algorithm}

Section 3.1 highlights quantum problem-solution correlation in Grover's
algorithm. In Sections 3.2 through 3.4, we develop the notion of advanced knowledge.

\subsection{Quantum problem-solution correlation}

We formalize the four drawer example. Bob chooses a two-bit number
$\mathbf{b}\equiv b_{0}b_{1}$; Alice should find the value of $\mathbf{b}$ by
computing the Kronecker function $\delta\left(  \mathbf{b},\mathbf{a}\right)
$ for various values of $\mathbf{a}\equiv a_{0}a_{1}$. The value of
$\delta\left(  \mathbf{b},\mathbf{a}\right)  $ is $1$ if $\mathbf{a}%
=\mathbf{b}$, $0$\ otherwise, which tells Alice whether the ball is in drawer
$\mathbf{a}$.

Usually, the value of $\mathbf{b}$ chosen by Bob is thought to be hard-wired
inside the black box that computes $\delta\left(  \mathbf{b},\mathbf{a}%
\right)  $. To highlight quantum correlation, we add to the usual description
of the quantum algorithm an imaginary quantum register $B$ that contains the
hard-wired value\footnote{We have taken the expression "imaginary register"
from Ref. $\left[  7\right]  $, which hilights the problem-solution symmetry
of Grover's and the phase estimation \ algorithms.}. In Section 3.3, this
imaginary register will serve to represent the usual quantum algorithm (with
the hard-wired value) "with respect" to the observer Alice in the sense of
relational quantum mechanics.

In view of this representation, we should consider the entire process of
determination of the value of $\mathbf{b}$. Thus, we assume that register $B$
is initially in a maximally mixed state. Register $A$ contains the value of
$\mathbf{a}$ tried by Alice, the one-qubit register $V$ (like "value") is
meant to contain the result of the computation of $\delta\left(
\mathbf{b},\mathbf{a}\right)  $, modulo $2$\ added to its former content for
logical reversibility. The two latter registers are initially prepared as
required by Grover's algorithm:%
\begin{align}
\left\vert \psi\right\rangle  &  =\frac{1}{4\sqrt{2}}\left(  \operatorname{e}%
^{i\varphi_{0}}\left\vert 00\right\rangle _{B}+\operatorname{e}^{i\varphi_{1}%
}\left\vert 01\right\rangle _{B}+\operatorname{e}^{i\varphi_{2}}\left\vert
10\right\rangle _{B}+\operatorname{e}^{i\varphi_{3}}\left\vert 11\right\rangle
_{B}\right) \nonumber\\
&  \left(  \left\vert 00\right\rangle _{A}+\left\vert 01\right\rangle
_{A}+\left\vert 10\right\rangle _{A}+\left\vert 11\right\rangle _{A}\right)
\left(  \left\vert 0\right\rangle _{V}-\left\vert 1\right\rangle _{V}\right)
. \label{one}%
\end{align}

To prepare $B$, Bob needs to know its state. To this end, he measures $\hat
{B}$ in state (\ref{one}). This randomly selects a value of $\mathbf{b}$, say
$01$, projecting (\ref{one}) on:%
\begin{equation}
P_{\alpha}\left\vert \psi\right\rangle =\frac{1}{2\sqrt{2}}\left\vert
01\right\rangle _{B}\left(  \left\vert 00\right\rangle _{A}+\left\vert
01\right\rangle _{A}+\left\vert 10\right\rangle _{A}+\left\vert
11\right\rangle _{A}\right)  \left(  \left\vert 0\right\rangle _{V}-\left\vert
1\right\rangle _{V}\right)  , \label{two}%
\end{equation}
we denote projection operators by the letter $P$. The two-bit entropy of the
quantum state goes to zero with the determination of the value of $\mathbf{b}$.

Then he applies to register $B$ a permutation of the values of $\mathbf{b}$ --
a unitary transformation $U_{B}$ -- that changes the randomly selected value
of $\mathbf{b}$ into the desired one, say $00$:%
\begin{equation}
U_{B}P_{\alpha}\left\vert \psi\right\rangle =\frac{1}{2\sqrt{2}}\left\vert
00\right\rangle _{B}\left(  \left\vert 00\right\rangle _{A}+\left\vert
01\right\rangle _{A}+\left\vert 10\right\rangle _{A}+\left\vert
11\right\rangle _{A}\right)  \left(  \left\vert 0\right\rangle _{V}-\left\vert
1\right\rangle _{V}\right)  . \label{three}%
\end{equation}
We note that Bob can choose the desired value off-line in any way, for example
random with whatever probability distribution.

The reversible computation of $\delta\left(  \mathbf{b},\mathbf{a}\right)  $,
represented by the unitary transformation $U_{f}$ ($f$ like "function
evaluation"), sends (\ref{three}) into:%

\begin{equation}
U_{f}U_{B}P_{\alpha}\left\vert \psi\right\rangle =\frac{1}{2\sqrt{2}%
}\left\vert 00\right\rangle _{B}\left(  -\left\vert 00\right\rangle
_{A}+\left\vert 01\right\rangle _{A}+\left\vert 10\right\rangle _{A}%
+\left\vert 11\right\rangle _{A}\right)  (\left\vert 0\right\rangle
_{V}-\left\vert 1\right\rangle _{V}). \label{secondstage}%
\end{equation}
A non-computational operation, the rotation $U_{A}$ of the\ measurement basis
of register $A$ (the so called "inversion about the mean") yields:%

\begin{equation}
U_{A}U_{f}U_{B}P_{\alpha}\left\vert \psi\right\rangle =\frac{1}{\sqrt{2}%
}\left\vert 00\right\rangle _{B}\left\vert 00\right\rangle _{A}\left(
\left\vert 0\right\rangle _{V}-\left\vert 1\right\rangle _{V}\right)  .
\label{four}%
\end{equation}
In (\ref{four}), register $A$ contains the solution of the problem -- the
value of $\mathbf{b}$ chosen by Bob. Alice acquires it by measuring $\hat{A}$,
which by the way leaves state (\ref{four}) unaltered.

Of course, there is a one-to-one correlation between the value of $\mathbf{b}$
chosen by Bob and the solution found by Alice. Up to the permutation
introduced by $U_{B}$, this corresponds to the quantum correlation between the
outcome of measuring $\hat{B}$ in (\ref{one}) and that of measuring $\hat{A}$
in (\ref{four}). As anticipated in the Extended summary, from the standpoint
of quantum correlation, which concerns repetitions of the same quantum
experiment, the permutation $U_{B}$ should be considered fixed (the fact that
Bob chooses it to obtain the desired value of $\mathbf{b}$ belongs to a
different film).

With a fixed $U_{B}$, all is like the value of $\mathbf{b}$ chosen by Bob was
randomly selected -- this value becomes the fixed permutation of the randomly
selected value. Moreover, we can virtually defer the measurement of $\hat{B}$
at the end of the algorithm (in Section 3.3, we will show that this yields the
quantum algorithm "relativized" to the observer Alice). The previous
input-output sequence\ of ket vectors becomes:%
\begin{align}
\left\vert \psi\right\rangle  &  =\frac{1}{4\sqrt{2}}\left(  \operatorname{e}%
^{i\varphi_{0}}\left\vert 00\right\rangle _{B}+\operatorname{e}^{i\varphi_{1}%
}\left\vert 01\right\rangle _{B}+\operatorname{e}^{i\varphi_{2}}\left\vert
10\right\rangle _{B}+\operatorname{e}^{i\varphi_{3}}\left\vert 11\right\rangle
_{B}\right) \nonumber\\
&  \left(  \left\vert 00\right\rangle _{A}+\left\vert 01\right\rangle
_{A}+\left\vert 10\right\rangle _{A}+\left\vert 11\right\rangle _{A}\right)
\left(  \left\vert 0\right\rangle _{V}-\left\vert 1\right\rangle _{V}\right)
, \label{oned}%
\end{align}%
\begin{equation}
U_{B}\left\vert \psi\right\rangle =\left\vert \psi\right\rangle \label{dued}%
\end{equation}

\begin{equation}
U_{f}U_{B}\left\vert \psi\right\rangle =\frac{1}{4\sqrt{2}}\left[
\begin{array}
[c]{c}%
\operatorname{e}^{i\varphi_{0}}\left\vert 00\right\rangle _{K}\left(
-\left\vert 00\right\rangle _{X}+\left\vert 01\right\rangle _{X}+\left\vert
10\right\rangle _{X}+\left\vert 11\right\rangle _{X}\right)  +\\
\operatorname{e}^{i\varphi_{1}}\left\vert 01\right\rangle _{K}\left(
\left\vert 00\right\rangle _{X}-\left\vert 01\right\rangle _{X}+\left\vert
10\right\rangle _{X}+\left\vert 11\right\rangle _{X}\right)  +\\
\operatorname{e}^{i\varphi_{2}}\left\vert 10\right\rangle _{K}\left(
\left\vert 00\right\rangle _{X}+\left\vert 01\right\rangle _{X}-\left\vert
10\right\rangle _{X}+\left\vert 11\right\rangle _{X}\right)  +\\
\operatorname{e}^{i\varphi_{3}}\left\vert 11\right\rangle _{K}\left(
\left\vert 00\right\rangle _{X}+\left\vert 01\right\rangle _{X}+\left\vert
10\right\rangle _{X}-\left\vert 11\right\rangle _{X}\right)
\end{array}
\right]  (\left\vert 0\right\rangle _{V}-\left\vert 1\right\rangle _{V}),
\label{secondstaged}%
\end{equation}%
\begin{align}
U_{A}U_{f}U_{B}\left\vert \psi\right\rangle  &  =\frac{1}{2\sqrt{2}}\left(
\operatorname{e}^{i\varphi_{0}}\left\vert 00\right\rangle _{B}\left\vert
00\right\rangle _{A}+\operatorname{e}^{i\varphi_{1}}\left\vert 01\right\rangle
_{B}\left\vert 01\right\rangle _{A}+\operatorname{e}^{i\varphi_{2}}\left\vert
10\right\rangle _{B}\left\vert 10\right\rangle _{A}+\operatorname{e}%
^{i\varphi_{3}}\left\vert 11\right\rangle _{B}\left\vert 11\right\rangle
_{A}\right) \nonumber\\
&  \left(  \left\vert 0\right\rangle _{V}-\left\vert 1\right\rangle
_{V}\right)  , \label{threed}%
\end{align}%
\begin{equation}
P_{\omega}U_{A}U_{f}U_{B}\left\vert \psi\right\rangle =\frac{1}{\sqrt{2}%
}\left\vert 00\right\rangle _{B}\left\vert 00\right\rangle _{A}\left(
\left\vert 0\right\rangle _{V}-\left\vert 1\right\rangle _{V}\right)  .
\label{fourd}%
\end{equation}

In sending (\ref{oned}) into (\ref{dued}), $U_{B}$\ should permute the
suffixes of the $\varphi_{i}$. We do not take this into account since it is
irrelevant, given that the $\varphi_{i}$ are independent random phases. The
computation of $\delta\left(  \mathbf{b},\mathbf{a}\right)  $,\ namely $U_{f}%
$, maximally entangles the content of register $B$\ with that of register $A$.
In (\ref{secondstaged}), four orthogonal states of $B$, each a value of
$\mathbf{b}$, are correlated with four orthogonal states of $A$, which means
that the information about the value of $\mathbf{b}$\ has propagated to
register $A$. The rotation of the\ measurement basis of $A$ makes this
information readable: entanglement also becomes correlation between
measurement outcomes. We can see why Bob's measurement can be virtually
deferred at the end: the projection of (\ref{threed}) on (\ref{fourd}), back
evolved by $U_{B}^{\dag}U_{f}^{\dag}U_{A}^{\dag}$, becomes the projection of
(\ref{one}) on (\ref{two}).

Thinking that all measurements are performed in the maximally entangled state
(\ref{threed}) makes it more clear that the value of $\mathbf{b}$ is randomly
selected by either Bob's or Alice's measurement; by the way, since $\hat{A}$
and $\hat{B}$\ commute, the order of these two measurements (which can also be
simultaneous) in state (\ref{threed}) is irrelevant. Either measurement
projects state (\ref{threed}) on the \textit{solution eigenstate}
(\ref{fourd}), where both registers contain the selected value of $\mathbf{b}%
$; correspondingly, the two-bit entropy of the quantum state goes to zero.

In view of what will follow, it is useful to introduce the reduced density
operator (in the random phase representation) of register $B$ in state
(\ref{threed}):
\begin{equation}
\rho_{B}=\frac{1}{2}\left(  \operatorname{e}^{i\varphi_{0}}\left\vert
00\right\rangle _{B}+\operatorname{e}^{i\varphi_{1}}\left\vert 01\right\rangle
_{B}+\operatorname{e}^{i\varphi_{2}}\left\vert 10\right\rangle _{B}%
+\operatorname{e}^{i\varphi_{3}}\left\vert 11\right\rangle _{B}\right)  .
\label{densityb}%
\end{equation}
Incidentally, we note that $\rho_{B}$\ is the same in states (\ref{oned})
through (\ref{threed}), up to the irrelevant permutation of random phases; the
algorithm is in fact the identity on the reduced density operator of the
control register. The projection of state (\ref{threed}) on the solution
eigenstate (\ref{fourd}) implies of course the projection of $\rho_{B}$\ on
$\left\vert 00\right\rangle _{B}$; we will also say "on $\mathbf{b}\in\left\{
00\right\}  $". By the way, considering also $\rho_{A}$\ would lead to
completely redundant considerations.

\subsection{Sharing the projection on Bob's choice}

Given that now the value of $\mathbf{b}$ can be determined by measuring either
$\hat{A}$ in state (\ref{threed})\ or $\hat{B}$ in the same or any former
state, we share the projection of state (\ref{threed}) on (\ref{fourd}) into
\textit{two projections}, one ascribed to Alice's measurement the other to
Bob's, in such a way that conditions (i) through (iii) of the sharing rule
described in the Extended summary are satisfied.

In the case of Grover's algorithm, the (in general) $n$\ bits that specify the
value of $\mathbf{b}$\ are independently selected in a random way (as the
fixed permutation of a similar selection). Thus, condition (i) states that the
projection on $p$\ of these bits ($0\leq p\leq n$) should be ascribed to
Alice's measurement, that on the other $n-p$ bits to Bob's. This also means
ascribing an entropy reduction of $p$ bits to Alice's measurements, of $n-p$
bits to Bob's. Thus, condition (ii) implies $p=n/2$.

As also anticipated in the Extended summary, the physical meaning of sharing
the projection on the value of $\mathbf{b}$ is related to the notion of
partial measurement of $\hat{B}$. For example, going back to $n=2$, we can
think of measuring $\hat{B}_{0}$, the content of the left cell of register
$B$, in state (\ref{threed}). This yields either $b_{0}=0$ or $b_{0}=1$,
projecting $\rho_{B}$ on either $\frac{1}{\sqrt{2}}\left(  \operatorname{e}%
^{i\varphi_{0}}\left\vert 00\right\rangle _{B}+\operatorname{e}^{i\varphi_{1}%
}\left\vert 01\right\rangle _{B}\right)  $ or, respectively, $\frac{1}%
{\sqrt{2}}\left(  \operatorname{e}^{i\varphi_{2}}\left\vert 10\right\rangle
_{B}+\operatorname{e}^{i\varphi_{3}}\left\vert 11\right\rangle _{B}\right)  $.
In the present assumption, the overall measurement of $\hat{B}$ projects
$\rho_{B}$ on $\mathbf{b}\in\left\{  00\right\}  $, we are in fact discussing
how to share this projection. This naturally implies the assumption that the
measurement of $\hat{B}_{0}$ projects $\rho_{B}$ on $\frac{1}{\sqrt{2}}\left(
\operatorname{e}^{i\varphi_{0}}\left\vert 00\right\rangle _{B}%
+\operatorname{e}^{i\varphi_{1}}\left\vert 01\right\rangle _{B}\right)  $; we
also say "on $\mathbf{b}\in\left\{  00,01\right\}  $". Under the same
assumption, measuring $\hat{B}_{1}$, the content of the right cell of $B$,
projects $\rho_{B}$ on $\mathbf{b}\in\left\{  00,10\right\}  $. There is still
one partial measurement that yields one bit of information about the value of
$\mathbf{b}$, that of $\hat{B}_{X}$, the exclusive or of the contents of the
two cells. Measuring $\hat{B}_{X}$ projects -- always under the same
assumption --\ on $\mathbf{b}\in\left\{  00,11\right\}  $. Summing up, a
possible way of dividing the projection of $\rho_{B}$ on $\mathbf{b}%
\in\left\{  00\right\}  $ is to split it into any two of the following three
projections, on: $\mathbf{b}\in\left\{  00,01\right\}  $, $\mathbf{b}%
\in\left\{  00,10\right\}  $, and $\mathbf{b}\in\left\{  00,11\right\}  $. In
the four drawers visualization, this means of course pairing drawer $00$\ with
another drawer in all possible ways.

This example is not fortuitous. In fact, it is the only way of sharing
(between Alice's and Bob's measurements) the projection on $\mathbf{b}%
\in\left\{  00\right\}  $ that satisfies conditions (i) and (ii). One can
readily see that the measurement of any pair of observables among $\hat{B}%
_{0}$, $\hat{B}_{1}$, and $\hat{B}_{X}$, selects a value of $\mathbf{b}$
without projecting twice on any bit of this value. Furthermore, the entropy
drop associated with either one of the two measurements is the same. One can
also see that there is no other way of satisfying the above conditions;
condition (iii) will be addressed further below.

We provide an example for $n=4$. The projection on, say, $\mathbf{b}%
\in\left\{  0000\right\}  $ can be shared, say, in the projection on
$\mathbf{b}\in\left\{  0000,0001,0010,0011\right\}  $ and that on
$\mathbf{b}\in\left\{  0000,0100,1000,1100\right\}  $. The former projection
corresponds to measuring $\hat{B}_{0}$ and $\hat{B}_{1}$\ and finding
$b_{0}=b_{1}=0$, the latter to measuring $\hat{B}_{2}$ and $\hat{B}_{3}$\ and
and finding $b_{2}=b_{3}=0$.

\subsection{Advanced knowledge}

We show that ascribing to Alice part of Bob's choice,\ implies that she knows
in advance, before running the algorithm, that part of the choice.

To this end, we introduce the notion of \textit{relativized} quantum
algorithm, in the sense of relational quantum mechanics $\left[  8\right]  $.
We note that states (\ref{dued}) through (\ref{fourd}) are the original
quantum algorithm -- we mean states (\ref{three}) through (\ref{four}) (to be
counted twice, before and after the measurement of $\hat{A}$) -- but with the
quantum state relativized to the observer Alice. By definition, initially
Alice does not know the content of register $B$. To her, register $B$ is in a
maximally mixed state even if Bob has already chosen the value of $\mathbf{b}%
$. The two-bit entropy of state (\ref{dued}) represents Alice's complete
ignorance of the value of $\mathbf{b}$. When Alice measures $\hat{A}$ at the
end of the algorithm, the quantum state (\ref{threed}) is projected on the
solution eigenstate (\ref{fourd}). This projection is random to Alice, it is
actually on the value of $\mathbf{b}$\ chosen by Bob. The entropy of the
quantum state goes to zero and Alice acquires full knowledge of the value of
$\mathbf{b}$. Thus, the entropy of the relativized quantum state gauges
Alice's ignorance of the value of $\mathbf{b}$ throughout the execution of the algorithm.

With this result, we go back to our aim. We work on an example. We share the
projection of state (\ref{threed}) on (\ref{fourd}), namely the selection of
$\mathbf{b}=00$, by ascribing to Alice's measurement the projection of
(\ref{threed}) on:%
\begin{equation}
\frac{1}{2}\left(  \operatorname{e}^{i\varphi_{0}}\left\vert 00\right\rangle
_{B}\left\vert 00\right\rangle _{A}+\operatorname{e}^{i\varphi_{1}}\left\vert
01\right\rangle _{B}\left\vert 01\right\rangle _{A}\right)  \left(  \left\vert
0\right\rangle _{V}-\left\vert 1\right\rangle _{V}\right)  , \label{pro}%
\end{equation}
namely the selection of $a_{0}=b_{0}=0$. This selection is like it was
randomly generated at the time and location of Alice's measurement. To become
a contribution to Bob's choice (itself the fixed permutation of a randomly
generated value), it must propagate to the time and location of Bob's choice
-- in particular to before running the algorithm and immediately after
applying $U_{B}$. Therefore, we should back evolve the corresponding
projection by applying $U_{B}^{\dag}U_{f}^{\dag}U_{A}^{\dag}$ to the two ends
of it, namely to states (\ref{threed}) and (\ref{pro}). This yields the
projection of the initial state (\ref{dued}) \ -- or, identically,
(\ref{oned}) -- on:%
\begin{equation}
\frac{1}{4}\left(  \operatorname{e}^{i\varphi_{0}}\left\vert 00\right\rangle
_{B}+\operatorname{e}^{i\varphi_{1}}\left\vert 01\right\rangle _{B}\right)
\left(  \left\vert 00\right\rangle _{A}+\left\vert 01\right\rangle
_{A}+\left\vert 10\right\rangle _{A}+\left\vert 11\right\rangle _{A}\right)
\left(  \left\vert 0\right\rangle _{V}-\left\vert 1\right\rangle _{V}\right)
.
\end{equation}
The entropy of the initial state of register $B$ is halved. Since this entropy
represents Alice's initial ignorance of Bob's choice, this means that Alice,
before running the algorithm, knows $n/2$ \ of the bits of Bob's choice, here
one bit -- in fact $b_{0}=0$.

We are at the level of elementary logical operations, where knowing means
doing. Alice knows half of Bob's choice (the value of $b_{0}$) by acting like
she knew it, namely by using it to identify classically the missing half (the
value of $b_{1}$) with a single computation of $\delta\left(  \mathbf{b}%
,\mathbf{a}\right)  $. Correspondingly, as we showed in $\left[  6\right]  $,
the quantum algorithm is the superposition of all the possible ways of taking
one bit of information about Bob's choice and, given the advanced knowledge of
this bit, classically identifying the missing bit with a single computation of
$\delta\left(  \mathbf{b},\mathbf{a}\right)  $ -- see also Section 3.4. This
explains the speed-up from three to one computation. This also satisfies
condition (iii) of our sharing rule, namely that the projection on the value
of $\mathbf{b}$ is shared between Alice's and Bob's measurements in all
possible ways (compatibly with the other conditions) in quantum superposition.

This explanation of the mechanism of the speed-up generalizes to $\mathbf{b}$
any number of bits -- see Ref. $\left[  6\right]  $.

\subsection{Superposition of classical computation histories}

We show in which sense the quantum algorithm can be seen as a superposition of
classical computations. As we have seen, in the assumption that Bob's choice
is $\mathbf{b}=00$, Alice's advanced knowledge can be: $\mathbf{b}\in\left\{
00,01\right\}  $, or $\mathbf{b}\in\left\{  00,10\right\}  $, or
$\mathbf{b}\in\left\{  00,11\right\}  $. We start with the first possibility.
Given the advanced knowledge of $\mathbf{b}\in\left\{  00,01\right\}  $, to
identify the value of $\mathbf{b}$ Alice should compute $\delta\left(
\mathbf{b},\mathbf{a}\right)  $ for either $\mathbf{a}=00$ or $\mathbf{a}=01$.
Let us assume it is for $\mathbf{a}=00$. The outcome of the computation is
$\delta=1$. This originates two classical computation \textit{histories},
depending on whether the initial state of register $V$ is $\left\vert
0\right\rangle _{V}$ or $\left\vert 1\right\rangle _{V}$. Each classical
history is represented as a sequence of sharp quantum states, as follows. The
initial state of history 1 is $\operatorname{e}^{i\varphi_{0}}\left\vert
00\right\rangle _{B}\left\vert 00\right\rangle _{A}\left\vert 0\right\rangle
_{V}$ (the ket $\left\vert 00\right\rangle _{B}$ means that $\mathbf{b}=00$,
the ket $\left\vert 00\right\rangle _{A}$ that the input of the computation of
$\delta$\ is $\mathbf{a}=00$); the state after the computation of $\delta$ is
$\operatorname{e}^{i\varphi_{0}}\left\vert 00\right\rangle _{B}\left\vert
00\right\rangle _{A}\left\vert 1\right\rangle _{V}$ (the result of the
computation is modulo 2 added to the former content of register $V$). We are
using the history phases that reconstruct the quantum algorithm: our present
aim is to show that the quantum algorithm is a superposition of classical
computation histories\footnote{History phases can also be found from scratch
by maximizing entanglement -- see Ref. $\left[  6\right]  $.}. In history 2,
the states before/after the computation of $\delta$ are $-\operatorname{e}%
^{i\varphi_{0}}\left\vert 00\right\rangle _{B}\left\vert 00\right\rangle
_{A}\left\vert 1\right\rangle _{V}\rightarrow-\operatorname{e}^{i\varphi_{0}%
}\left\vert 00\right\rangle _{B}\left\vert 00\right\rangle _{A}\left\vert
0\right\rangle _{V}$. In the case that Alice computes $\delta\left(
\mathbf{b},\mathbf{a}\right)  $ for $\mathbf{a}=01$ instead, she obtains
$\delta=0$, which of course tells her again that $\mathbf{b}=00$. This
originates other two histories. History 3: $\operatorname{e}^{i\varphi_{0}%
}\left\vert 00\right\rangle _{B}\left\vert 01\right\rangle _{A}\left\vert
0\right\rangle _{V}\rightarrow\operatorname{e}^{i\varphi_{0}}\left\vert
00\right\rangle _{B}\left\vert 01\right\rangle _{A}\left\vert 0\right\rangle
_{V}$; history 4: $-\operatorname{e}^{i\varphi_{0}}\left\vert 00\right\rangle
_{B}\left\vert 01\right\rangle _{A}\left\vert 1\right\rangle _{V}%
\rightarrow-\operatorname{e}^{i\varphi_{0}}\left\vert 00\right\rangle
_{B}\left\vert 01\right\rangle _{A}\left\vert 1\right\rangle _{V}$. We develop
in a similar way the other possibilities, also for all the possible choices of
the value of $\mathbf{b}$. The computation step of Grover's algorithm, namely
the transformation of (\ref{dued}) into (\ref{secondstaged}), is the
superposition of all these histories.

By the way, this also explains quantum parallel computation. In fact, in the
initial state of the quantum algorithm and in the superposition of all
classical computation histories, each $\left\vert ij\right\rangle _{B}$ is
multiplied by:%
\begin{equation}
\frac{1}{2\sqrt{2}}\left(  \left\vert 00\right\rangle _{A}+\left\vert
01\right\rangle _{A}+\left\vert 10\right\rangle _{A}+\left\vert
11\right\rangle _{A}\right)  \left(  \left\vert 0\right\rangle _{V}-\left\vert
1\right\rangle _{V}\right)  ,
\end{equation}
as one can readily check.

As one can see, the computation step we are dealing with is the identity on
the reduced density operator of register $B$ (the control register) and
entangles this register with register $A$ (the target register). The
information contained in $B$ leaks to $A$ -- see state (\ref{secondstaged}).
At this point we perform a non-computational step: a suitable rotation $U_{A}$
of the basis of register $A$ (the so called "inversion about the mean"). This
branches each history into four histories; such branches interfere with one
another to give state (\ref{threed}). Entanglement also becomes correlation
between the possible measurement outcomes. By the way, this implicitly defines
$U_{A}$ (inversion about the mean) as the rotation of the basis of register
$A$\ that maximizes correlation between possible measurement outcomes.

\ Summing up, Grover's algorithm can be decomposed into a superposition of
histories, which start from Alice's advanced knowledge and whose computational
part is entirely classical. This result also applies to $n>2$, by iterating
the sequence of the two steps (computational and non computational)
$\operatorname{O}\left(  2^{n/2}\right)  $ times. By the way, this means a
"quadratic" speed-up with respect to a classical algorithm that requires
$\operatorname{O}\left(  2^{n}\right)  $ computations.

\section{Deutsch\&Jozsa's algorithm}

In Deutsch\&Jozsa's $\left[  9\right]  $ algorithm, the set of functions known
to both Bob and Alice is all the constant and "balanced" functions (with an
even number of zeroes and ones) $f_{\mathbf{b}}:\left\{  0,1\right\}
^{n}\rightarrow\left\{  0,1\right\}  $. Array (\ref{dj}) gives this set for
$n=2$. The string $\mathbf{b}\equiv b_{0},b_{1},...,b_{2^{n}-1}$ is both the
suffix and the table of the function -- the sequence of function values for
increasing values of the argument. Specifying the choice of the function by
means of the table of the function simplifies the discussion.
\begin{equation}%
\begin{tabular}
[c]{|c|c|c|c|c|c|c|c|c|}\hline
$\mathbf{a}$ & $\,f_{0000}\left(  \mathbf{a}\right)  $ & $f_{1111}\left(
\mathbf{a}\right)  $ & $f_{0011}\left(  \mathbf{a}\right)  $ & $f_{1100}%
\left(  \mathbf{a}\right)  $ & $f_{0101}\left(  \mathbf{a}\right)  $ &
$f_{1010}\left(  \mathbf{a}\right)  $ & $f_{0110}\left(  \mathbf{a}\right)  $
& $f_{1001}\left(  \mathbf{a}\right)  $\\\hline
00 & 0 & 1 & 0 & 1 & 0 & 1 & 0 & 1\\\hline
01 & 0 & 1 & 0 & 1 & 1 & 0 & 1 & 0\\\hline
10 & 0 & 1 & 1 & 0 & 0 & 1 & 1 & 0\\\hline
11 & 0 & 1 & 1 & 0 & 1 & 0 & 0 & 1\\\hline
\end{tabular}
\ \ \label{dj}%
\end{equation}

Alice should find whether the function selected by Bob is balanced or constant
by computing $f_{\mathbf{b}}\left(  \mathbf{a}\right)  \equiv f\left(
\mathbf{b},\mathbf{a}\right)  $ for appropriate values of $\mathbf{a}$. In the
classical case this requires, in the worst case, a number of computations of
$f\left(  \mathbf{b},\mathbf{a}\right)  $ exponential in $n$; in the quantum
case one computation.

We give the relativized states before and after the unitary part of the
algorithm, namely $U_{A}U_{f}$ ($U_{B}$, $U_{f}$, and $U_{A}$ play the same
role as before but are of course specific to the algorithm):%
\begin{align}
U_{B}\left\vert \psi\right\rangle  &  =\frac{1}{8}\left(  \operatorname{e}%
^{i\varphi_{0}}\left\vert 0000\right\rangle _{B}+\operatorname{e}%
^{i\varphi_{1}}\left\vert 1111\right\rangle _{B}+\operatorname{e}%
^{i\varphi_{2}}\left\vert 0011\right\rangle _{B}+\operatorname{e}%
^{i\varphi_{3}}\left\vert 1100\right\rangle _{B}+~...\right) \nonumber\\
&  \left(  \left\vert 00\right\rangle _{A}+\left\vert 01\right\rangle
_{A}+\left\vert 10\right\rangle _{A}+\left\vert 11\right\rangle _{A}\right)
\left(  \left\vert 0\right\rangle _{V}-\left\vert 1\right\rangle _{V}\right)
\label{twodj}%
\end{align}%
\begin{align}
U_{A}U_{f}U_{B}\left\vert \psi\right\rangle  &  =\frac{1}{4}\left[  \left(
\operatorname{e}^{i\varphi_{0}}\left\vert 0000\right\rangle _{B}%
-\operatorname{e}^{i\varphi_{1}}\left\vert 1111\right\rangle _{B}\right)
\left\vert 00\right\rangle _{A}+\left(  \operatorname{e}^{i\varphi_{2}%
}\left\vert 0011\right\rangle _{B}-\operatorname{e}^{i\varphi_{3}}\left\vert
1100\right\rangle _{B}\right)  \left\vert 10\right\rangle _{A}+~...\right]
\nonumber\\
&  \left(  \left\vert 0\right\rangle _{V}-\left\vert 1\right\rangle
_{V}\right)  . \label{threedj}%
\end{align}
The entangled state (\ref{threedj}) is reached with a single computation of
$f\left(  \mathbf{b},\mathbf{a}\right)  $. Measuring $\hat{A}$\ in
(\ref{threedj}) yields the solution: all zeros if the function is constant,
not so if it is balanced.

This time entanglement is a-symmetric and we should consider the reduced
density operators of both registers $B$ and $A$ in state (\ref{threedj}):
\begin{equation}
\rho_{B}=\frac{1}{2\sqrt{2}}\left(  \operatorname{e}^{i\varphi_{0}}\left\vert
0000\right\rangle _{B}+\operatorname{e}^{i\varphi_{1}}\left\vert
1111\right\rangle _{B}+\operatorname{e}^{i\varphi_{2}}\left\vert
0011\right\rangle _{B}+\operatorname{e}^{i\varphi_{3}}\left\vert
1100\right\rangle _{B}+~...\right)  , \label{roa}%
\end{equation}%
\begin{equation}
\rho_{A}=\frac{1}{2}\left(  \operatorname{e}^{i\vartheta_{0}}\left\vert
00\right\rangle _{A}+\operatorname{e}^{i\vartheta_{1}}\left\vert
01\right\rangle _{A}+\operatorname{e}^{\vartheta_{2}}\left\vert
10\right\rangle _{B}+\operatorname{e}^{i\vartheta_{3}}\left\vert
11\right\rangle _{A}\right)  , \label{rob}%
\end{equation}
the $\vartheta_{i}$\ are independent random phases with uniform distribution
in $\left[  0,2\pi\right]  $ as well. This time the entropies of $\rho_{B}$
and $\rho_{A}$ are different, $3$ bits and $2$\ bits respectively.
Incidentally, we note that $\rho_{B}$\ remains unaltered throughout the
unitary transformation $U_{A}U_{f}U_{B}$.

We lay down the two lion's share perspectives. If we assume that the
measurement of $\hat{B}$ is performed first, we ascribe to it: the zeroing of
the entropy of $\rho_{B}$ and the zeroing of the entropy of $\rho_{A}$. If we
assume that the measurement of $\hat{A}$ is performed first, we ascribe to it:
the zeroing of the entropy of $\rho_{A}$ and the reduction of the entropy of
$\rho_{B}$.

Condition (ii) of the sharing rule becomes that each one of "the two
projections" (as defined in the sharing rule) properly reduces both entropies.
This is enough to univocally solve the sharing problem.

To perform the sharing, in the first place we should identify the "elementary"
partial projections. A natural choice is considering the projections
associated with measuring $\hat{B}_{0}$, $\hat{B}_{1}$, ... , of course
selecting the measurement outcomes that match with Bob's choice. To fix ideas,
we assume that Bob's choice is $\mathbf{b}=0011$. This means considering the
projections on the single bits of the bit string $\mathbf{b}=0011$ or, in
equivalent terms, on the single rows of the table of $f_{\mathbf{b}}$ -- see
the third column of array (\ref{dj}). As we will see, this provides sufficient
resolution to build the history superposition picture of the quantum
algorithm; considering also Boolean functions of the bit string $0011$ would
generate repeated histories here.

Summing up, we should select two shares of the table of the function (the
projection on one should be ascribed to Alice's measurement, that on the other
to Bob's) in such a way that, together, they project on the value of
$\mathbf{b}$ without over-projecting on any part of it and, individually,
reduce the entropies of both $\rho_{B}$ and $\rho_{A}$.

This implies that no share contains different values of the function or, if
all the values are the same, more than 50\% of the rows. Otherwise, the
projection on that share would already tell that the function is balanced, or
constant. For the no over-projection condition, this would mean ascribing to
only one projection the entire reduction of the entropy of $\rho_{A}$, against
condition (ii).

Conditions (i) and (ii) are satisfied if the two shares of Bob's choice
$\mathbf{b}=0011$ are respectively $f_{\mathbf{b}}\left(  00\right)
=0,f_{\mathbf{b}}\left(  01\right)  =0$ and $f_{\mathbf{b}}\left(  10\right)
=1,f_{\mathbf{b}}\left(  11\right)  =1$ -- see array (\ref{dj}). One can
easily check that any deviation from this sharing violates the aforesaid
conditions. For example, if the two shares were instead $f_{\mathbf{b}}\left(
00\right)  =0$ and $f_{\mathbf{b}}\left(  11\right)  =1$, projecting on them
would not determine Bob's choice, thus violating condition (i). If they were
$f_{\mathbf{b}}\left(  00\right)  =0,f_{\mathbf{b}}\left(  01\right)  =0$ and
$f_{\mathbf{b}}\left(  11\right)  =1$, this would determine Bob's choice, but
the projection on the latter share would not reduce the entropy of $\rho_{A}$,
thus violating condition (ii). Etc. We call either one of the two shares of
the table a \textit{good half table}.

Incidentally, nothing a-priori requires that we split the entire table of the
function into two shares. In the quantum part of Shor's $\left[  10\right]  $
factorization algorithm -- finding the period $R$ of a periodic function --
conditions (i) and (ii) dictate that one share of the table is a set of $R$
consecutive rows, the other share a similar set with arguments displaced by a
multiple of $R$\ (the two sets should be taken in all possible ways in quantum
superposition). Splitting the entire table into two shares, if the domain of
the function spans more than two periods, would imply over-projection.

Back to Deutsch and Jozsa's algorithm, besides Alice's contribution to Bob's
choice, a good half table represents Alice's advanced knowledge of this
choice. In fact, since $\rho_{B}$ remains unaltered throughout the unitary
part of the quantum algorithm, also the projection of $\rho_{B}$ on a good
half table (on the superposition of the values of $\mathbf{b}$\ that match
with it) remains unaltered. At the end of the relativized quantum algorithm,
this projection represents Alice's contribution to Bob's choice. At the
beginning, it changes Alice's complete ignorance of Bob's choice into
knowledge of the good half table.

It is immediate to check that the quantum algorithm requires the number of
function evaluations of a classical algorithm that knows in advance a good
half table. In fact, the value of $\mathbf{b}$,\ and thus the solution, are
always identified by computing $f_{\mathbf{b}}\left(  \mathbf{a}\right)  $ for
only one value of $\mathbf{a}$\ (anyone) outside the half table -- see array
(\ref{dj}). Thus, both the quantum algorithm and the advanced knowledge
classical algorithm require just one function evaluation.

Now we go to the history superposition picture. It is convenient to group the
histories with the same value of $\mathbf{b}$. Starting with $\mathbf{b}%
=0011$, we assume that Alice's advanced knowledge is the good half table
$f\left(  \mathbf{b},00\right)  =0,f\left(  \mathbf{b},01\right)  =0$. As this
is common to $\mathbf{b}=0000$ and $\mathbf{b}=0011$, in order to find the
value of $\mathbf{b}$ and thus the character of the function, Alice should
perform function evaluation for either $\mathbf{a}=10$ or $\mathbf{a}=11$. We
assume it is for $\mathbf{a}=10$. Since we are under the assumption
$\mathbf{b}=0011$, the result of the computation is $1$. This originates two
classical computation histories, each consisting of a state before and one
after function evaluation. History 1: $\operatorname{e}^{i\varphi_{2}%
}\left\vert 0011\right\rangle _{B}\left\vert 10\right\rangle _{A}\left\vert
0\right\rangle _{V}\rightarrow\operatorname{e}^{i\varphi_{2}}\left\vert
0011\right\rangle _{B}\left\vert 10\right\rangle _{A}\left\vert 1\right\rangle
_{V}$; history 2: $-\operatorname{e}^{i\varphi_{2}}\left\vert
0011\right\rangle _{B}\left\vert 10\right\rangle _{A}\left\vert 1\right\rangle
_{V}\rightarrow-\operatorname{e}^{i\varphi_{2}}\left\vert 0011\right\rangle
_{B}\left\vert 10\right\rangle _{A}\left\vert 0\right\rangle _{V}$.\ If she
performs function evaluation for $\mathbf{a}=11$ instead, this originates
other two histories, etc.

In this way, in the state before function evaluation, the value of
$\mathbf{b}$\ is multiplied by the superposition of all the possible
combinations of values of $\mathbf{a}$ and contents of register $V$. This
explains quantum parallel computation. Performing a single computation of
$f\left(  \mathbf{b},\mathbf{a}\right)  $ entangles registers $B$ and $A$.

Until now we have seen the\ function evaluation stage of the quantum
algorithm. Before going further, it can be useful to summarize the overall
picture. Alice knows in advance half of the value of $\mathbf{b}$. In order to
find the solution, a function of $\mathbf{b}$, she should identify the entire
value of $\mathbf{b}$, by performing function evaluation for only one value of
$\mathbf{a}$\ (anyone) outside the half table. This is done in all possible
ways in quantum superposition. Function evaluation is the identity on
$\rho_{B}$; it entangles registers $B$ and $A$. Information contained in the
control register $B$ leaks to the target register $A$. Eventually, applying
the Hadamard transform to register $A$ yields state (\ref{threedj}):
entanglement also becomes correlation between the possible measurement outcomes.

By the way, the fact that Alice, in each individual history, knows half of the
value of $\mathbf{b}$ and computes the missing half in order to find the
solution, agrees with the fact that Alice cannot know the precise value of
$\mathbf{b}$\ by measuring $\hat{A}$\ in state (\ref{threedj}). In fact this
depends on the special form of state (\ref{threedj}), where each eigenstate of
$\hat{A}$ multiplies the superposition of two eigenstates of $\hat{B}$; this
form emerges in the very superposition of the individual histories.

It is easy to see that the present analysis, like the notion of
good-half-table, holds unaltered for $n>2$.

\section{Simon's and the hidden subgroup algorithms}

In Simon's $\left[  11\right]  $ algorithm, the set of functions is all the
$f_{\mathbf{b}}:\left\{  0,1\right\}  ^{n}\rightarrow\left\{  0,1\right\}
^{n-1}$ such that $f_{\mathbf{b}}\left(  \mathbf{a}\right)  =f_{\mathbf{b}%
}\left(  \mathbf{c}\right)  $ if and only if $\mathbf{a}=\mathbf{c}$\ or
$\mathbf{a}=\mathbf{c}\oplus\mathbf{h}^{\left(  \mathbf{b}\right)  }$;
$\oplus$\ denotes bitwise modulo 2 addition; the bit string $\mathbf{h}%
^{\left(  \mathbf{b}\right)  }$\textbf{, }depending on $\mathbf{b}$ and
belonging to $\left\{  0,1\right\}  ^{n}$ excluded the all zeroes string, is a
sort of period of the function. Array (\ref{periodic}) gives the set of
functions for $n=2$. The bit string $\mathbf{b}$ is both the suffix and the
table of the function. Since $\mathbf{h}^{\left(  \mathbf{b}\right)  }%
\oplus\mathbf{h}^{\left(  \mathbf{b}\right)  }=0$, each value of the function
appears exactly twice in the table, thus 50\% of the rows plus one surely
identify $\mathbf{h}^{\left(  \mathbf{b}\right)  }$.
\begin{equation}%
\begin{tabular}
[c]{|c|c|c|c|c|c|c|}\hline
& $\mathbf{h}^{\left(  0011\right)  }=01$ & $\mathbf{h}^{\left(  1100\right)
}=01$ & $\mathbf{h}^{\left(  0101\right)  }=10$ & $\mathbf{h}^{\left(
1010\right)  }=10$ & $\mathbf{h}^{\left(  0110\right)  }=11$ & $\mathbf{h}%
^{\left(  1001\right)  }=11$\\\hline
$\mathbf{a}$ & $f_{0011}\left(  \mathbf{a}\right)  $ & $f_{1100}\left(
\mathbf{a}\right)  $ & $f_{0101}\left(  \mathbf{a}\right)  $ & $f_{1010}%
\left(  \mathbf{a}\right)  $ & $f_{0110}\left(  \mathbf{a}\right)  $ &
$f_{1001}\left(  \mathbf{a}\right)  $\\\hline
00 & 0 & 1 & 0 & 1 & 0 & 1\\\hline
01 & 0 & 1 & 1 & 0 & 1 & 0\\\hline
10 & 1 & 0 & 0 & 1 & 1 & 0\\\hline
11 & 1 & 0 & 1 & 0 & 0 & 1\\\hline
\end{tabular}
\ \ \label{periodic}%
\end{equation}

Bob selects a value of $\mathbf{b}$. Alice's problem is finding the value of
$\mathbf{h}^{\left(  \mathbf{b}\right)  }$, "hidden" in $f_{\mathbf{b}}\left(
\mathbf{a}\right)  $, by computing $f_{\mathbf{b}}\left(  \mathbf{a}\right)
=f\left(  \mathbf{b},\mathbf{a}\right)  $ for different values of $\mathbf{a}%
$. In present knowledge, a classical algorithm requires a number of
computations of $f\left(  \mathbf{b},\mathbf{a}\right)  $ exponential in $n$.
The quantum algorithm solves the hard part of this problem, namely finding a
string $\mathbf{s}_{j}^{\left(  \mathbf{b}\right)  }$ orthogonal\footnote{The
modulo 2 addition of the bits of the bitwise product of the two strings should
be zero.} to $\mathbf{h}^{\left(  \mathbf{b}\right)  }$, with one computation
of $f\left(  \mathbf{b},\mathbf{a}\right)  $. There are $2^{n-1}$ such
strings. Running the quantum algorithm yields one of these strings at random
(see further below). The quantum algorithm is iterated until finding $n-1$
different strings. This allows us to find $\mathbf{h}^{\left(  \mathbf{b}%
\right)  }$ by solving a system of modulo 2 linear equations.

We give the relativized states before and after the unitary part of the algorithm:%

\begin{align}
U_{B}\left\vert \psi\right\rangle  &  =\frac{1}{2\sqrt{6}}\left(
\operatorname{e}^{i\varphi_{0}}\left\vert 0011\right\rangle _{B}%
+\operatorname{e}^{i\varphi_{1}}\left\vert 1100\right\rangle _{B}%
+\operatorname{e}^{i\varphi_{2}}\left\vert 0101\right\rangle _{B}%
+\operatorname{e}^{i\varphi_{3}}\left\vert 1010\right\rangle _{B}+~...\right)
\nonumber\\
&  \left(  \left\vert 00\right\rangle _{A}+\left\vert 01\right\rangle
_{A}+\left\vert 10\right\rangle _{A}+\left\vert 11\right\rangle _{A}\right)
\left\vert 0\right\rangle _{V}. \label{twos}%
\end{align}%
\begin{equation}
U_{A}U_{f}U_{B}\left\vert \psi\right\rangle =\frac{1}{2\sqrt{6}}\left\{
\begin{array}
[c]{c}%
(\operatorname{e}^{i\varphi_{0}}\left\vert 0011\right\rangle _{B}%
+\operatorname{e}^{i\varphi_{1}}\left\vert 1100\right\rangle _{B})\left[
(\left\vert 00\right\rangle _{A}+\left\vert 10\right\rangle _{A})\left\vert
0\right\rangle _{V}+(\left\vert 00\right\rangle _{A}-\left\vert
10\right\rangle _{A})\left\vert 1\right\rangle _{V}\right] \\
+(\operatorname{e}^{i\varphi_{2}}\left\vert 0101\right\rangle _{B}%
+\operatorname{e}^{i\varphi_{3}}\left\vert 1010\right\rangle _{B})\left[
(\left\vert 00\right\rangle _{A}+\left\vert 01\right\rangle _{A})\left\vert
0\right\rangle _{V}+(\left\vert 00\right\rangle _{A}-\left\vert
01\right\rangle _{A})\left\vert 1\right\rangle _{V}\right]  +~...
\end{array}
\right\}  . \label{threes}%
\end{equation}

\ In (\ref{twos}), register $V$ is prepared in the all zeros string (just one
zero for $n=2$). State (\ref{threes}) is reached with a single computation of
$f\left(  \mathbf{b},\mathbf{a}\right)  $. In (\ref{threes}), for each value
of $\mathbf{b}$, register $A$ (no matter the content of $V$) hosts even
weighted\ superpositions of the $2^{n-1}$ strings $\mathbf{s}_{j}^{\left(
\mathbf{b}\right)  }$ orthogonal to $\mathbf{h}^{\left(  \mathbf{b}\right)  }%
$. By measuring $\hat{A}$ in this state, Alice obtains at random one of the
$\mathbf{s}_{j}^{\left(  \mathbf{b}\right)  }$. Then we iterate the "right
part" of the algorithm (preparation of registers $A$\ and $V$, computation of
$f\left(  \mathbf{b},\mathbf{a}\right)  $, and measurement of $\hat{A}$) until
obtaining $n-1$ different $\mathbf{s}_{j}^{\left(  \mathbf{b}\right)  }$. \ 

Let $\rho_{B}$ and $\rho_{A}$ be the reduced density operators of respectively
$B$ and $A$ in state (\ref{threes}). We lay down the two lion's share
perspectives in the assumption that we always throw away the projections on
$\left\vert 00\right\rangle _{A}$, which do not reduce the entropy of either
$\rho_{B}$ or $\rho_{A}$\ (tell nothing about the value of $\mathbf{b}$ or the
solution). If we assume that the measurement of $\hat{B}$ is performed first,
we ascribe to it: the zeroing of the entropy of $\rho_{B}$ and the reduction
(zeroing in the case $n=2$) of the entropy of $\rho_{A}$. If we assume that
the measurement of $\hat{A}$ is performed first, we ascribe to it: the zeroing
of the entropy of $\rho_{A}$ and the reduction of the entropy of $\rho_{B}$.
This can readily be checked by looking at the form of state (\ref{threes}).

Condition (ii) of the sharing rule becomes that each one of "the two
projections" properly reduces both entropies. The analysis of the former
section -- about how to share the projection on the value of $\mathbf{b}$
between Alice's and Bob's measurements -- still holds. Now half of Bob's
choice is any half table that does not contain the same value of the function
twice, which would already specify the value of $\mathbf{h}^{\left(
\mathbf{b}\right)  }$ and thus of all the $\mathbf{s}_{j}^{\left(
\mathbf{b}\right)  }$.

We check that the quantum algorithm requires the number of function
evaluations of a classical algorithm that knows in advance a good half table.
In fact, the solution is always identified by computing $f\left(
\mathbf{b},\mathbf{a}\right)  $ for only one value of $\mathbf{a}$\ (anyone)
outside the half table. The new value of the function is necessarily a value
already present in the half table, which identifies $\mathbf{h}^{\left(
\mathbf{b}\right)  }$ and thus all the $\mathbf{s}_{j}^{\left(  \mathbf{b}%
\right)  }$. Thus, both the quantum algorithm and the advanced knowledge
classical algorithm require just one function evaluation.

We go to the history superposition picture. Let us assume that Bob choose
$\mathbf{b}=0011$. Alice's advanced knowledge is either $f\left(
\mathbf{b},01\right)  =0,f\left(  \mathbf{b},10\right)  =1$ or $f\left(
\mathbf{b},00\right)  =0,f\left(  \mathbf{b},11\right)  =1$. Let us start with
the former possibility. As this half table is common to $\mathbf{b}=0011$ and
$\mathbf{b}=1010$, in order to find the value of $\mathbf{b}$\ and thus the
character of the function, Alice should perform function evaluation for either
$\mathbf{a}=00$ or $\mathbf{a}=11$. We assume that it is for $\mathbf{a}=00$.
The result of the computation is $0$. This originates two classical
computation histories, each consisting of two states, before and after
function evaluation. History 1: $\operatorname{e}^{i\varphi_{0}}\left\vert
0011\right\rangle _{B}\left\vert 00\right\rangle _{A}\left\vert 0\right\rangle
_{V}\rightarrow\operatorname{e}^{i\varphi_{0}}\left\vert 0011\right\rangle
_{B}\left\vert 00\right\rangle _{A}\left\vert 0\right\rangle _{V}$; history 2:
$-\operatorname{e}^{i\varphi_{0}}\left\vert 0011\right\rangle _{B}\left\vert
00\right\rangle _{A}\left\vert 1\right\rangle _{V}\rightarrow-\operatorname{e}%
^{i\varphi_{0}}\left\vert 0011\right\rangle _{B}\left\vert 00\right\rangle
_{A}\left\vert 1\right\rangle _{V}$.\ If she performs function evaluation for
$\mathbf{a}=11$ instead, the result of the computation is $1$. This originates
other two histories, etc. The sum of all histories is the function evaluation
stage of the quantum algorithm. After function evaluation, we should apply the
Hadamard transform to register $A$. Each history branches into four histories;
branches interfere with one another to yield state (\ref{threes}).

The present analysis holds unaltered for $n>2$. It also applies to the
generalized Simon's problem and to the Abelian hidden subgroup problem. In
fact the corresponding algorithms are essentially the same as the algorithm
that solves Simon's problem. In the hidden subgroup problem, the set of
functions $f_{\mathbf{b}}:G\rightarrow W$ map a group $G$ to some finite set
$W$\ with the property that there exists some subgroup $S\leq G$ such that for
any $\mathbf{a},\mathbf{c}\in G$, $f_{\mathbf{b}}\left(  \mathbf{a}\right)
=f_{\mathbf{b}}\left(  \mathbf{c}\right)  $ if and only if $\mathbf{a}%
+S=\mathbf{c}+S$. The problem is to find the hidden subgroup $S$ by computing
$f_{\mathbf{b}}\left(  \mathbf{a}\right)  $ for various values of $\mathbf{a}%
$. Now, a large variety of problems solvable with a quantum speed-up can be
re-formulated in terms of the hidden subgroup problem $\left[  12,13\right]
$. Among these we find: the seminal Deutsch's problem, finding orders, finding
the period of a function (thus the problem solved by the quantum part of
Shor's factorization algorithm), discrete logarithms in any group, hidden
linear functions, self shift equivalent polynomials, Abelian stabilizer
problem, graph automorphism problem.

\section{Summary and conclusions}

We summarize and position the results obtained. The present explanation of the speed-up:

(i) Relies on solving a fundamental measurement problem: how to share between
redundant measurements the determination of correlated eigenvalues.

(ii) Shows that the apparent determinism of quantum algorithms is a visual illusion.

(iii) Applies to both quadratic and exponential speed-ups. The unifications
achieved so far, which focus on the unitary part of the quantum algorithm, do
not capture both kinds of speed-up.

(iv) Highlights the existence of a common scheme for the family of quantum
algorithms examined: (a) Given the advanced knowledge of half of Bob's choice,
\ Alice finds the missing half through function evaluations; each function
evaluation is the identity on the reduced density operator of the control
register $B$ and increases the entanglement between this register and the
target register $A$; information about Bob's choice flows from the former to
the latter register -- this is of course a "character" of the function chosen
by Bob. (b) By changing the basis of the target register after each function
evaluation, entanglement also becomes correlation between the possible
measurement outcomes (the application, after function evaluation, of the
"inversion about the mean", or the Hadamard transform, or the quantum Fourier
transform in the case of Shor's algorithm, maximizes this correlation). (c)
Steps (a) and (b) should be repeated the number of times required to
classically find the missing half of Bob's choice.

(v)\ Allows us to derive new quantum algorithms. Given a set of functions, we
go through steps (a), (b), and (c), and see what is the character of the
function obtained. Then we design the problem (finding that character of the
function) around this result. Ref. $\left[  6\right]  $ provides an example in point.

(vi)\ Provides a tool for ascertaining the quantum speed-up achievable in
solving a problem -- a central issue in quantum computation. The speed-up
comes from comparing two classical algorithms, with and without advanced
knowledge of half of Bob's choice.

(vii)\ Highlights the existence of special causality loops. Each individual
history contains such a loop: Alice knowing in advance half of Bob's choice
without computing it. This is possible because the missing computation is
performed in other histories and quantum interference provides\ cross-talk
between histories. The causality loop remains fully there in the superposition
of all histories.

Possible future work is trying and extend the present explanation to other
quantum algorithms and further investigating what the explanation means at a
fundamental physical level. One could expect cross fertilization between these
two prospects.

\subsection*{Acknowledgments}

Thanks are due to Vint Cerf, David Deutsch, Jens Eisert, Lov Grover, Tom
Toffoli, and Lev Vaidman for encouragement and comments; to Pablo Arrighi,
Artur Ekert, David Finkelstein, Hatmut Neven, and Daniel Sheehan for
encouragement and discussions.

\subsection*{References}

$\left[  1\right]  $ D. Deutsch, Proc. Roy. Soc. London; Series A,
Mathematical and Physical sciences \textbf{400}, 97 (1985).

$\left[  2\right]  $ L. K. Grover, in \textit{Proceedings of the 28th Annual
ACM Symposium on the Theory of Computing, Philadelphia, PA, May 22-24, 1996}
(ACM Press New York, 1996), p. 212.

$\left[  3\right]  $\ L. K. Grover, \textit{From Schr\"{o}dinger equation to
quantum search algorithm}, arXiv: quant-ph/0109116 (2001).

$\left[  4\right]  $\ D. Gross, S. T. Flammia, and J. Eisert, Phys. Rev. Lett.
\textbf{102} (19) (2009).

$\left[  5\right]  $ S. Hawking, \textit{On the Shoulders of Giants} (Running
Press, Philadelphia-London. ISBN 076241698x, 2003), p. 731.

$\left[  6\right]  $ G. Castagnoli, Phys. Rev. A, \textbf{82}, 052334 (2010)

$\left[  7\right]  $ F. Morikoshi, Int. J. Theor. Phys, DOI:
10.1007/s10773-011-0701-6 (2011).

$\left[  8\right]  $\ C. Rovelli, Int. J. Theor. Phys. \textbf{35}, 8, 1637 (1996).

$\left[  9\right]  $ D. Deutsch and R. Jozsa, Proc. R. Soc. London A,
\textbf{439}, 553 (1992).

$\left[  10\right]  $ P. W. Shor , in \textit{Proceedings of the 35th Annual
Symposium on the Foundations of Computer Science}, (IEEE Computer Society
Press, Los Alamitos, CA, 1994), p. 124.

$\left[  11\right]  $ D. Simon, in \textit{Proceedings of the 35th Annual
Symposium on the Foundations of Computer Science}, (IEEE Computer Society
Press, Los Alamitos, CA, 1994), p.116.

$\left[  12\right]  $ M. Mosca, A. Ekert, in "Quantum Computing and Quantum
Communications", Volume 1509, Issue, May, Springer, p. 16 (1999).

$\left[  13\right]  $ P. Kaye, R. Laflamme, M. Mosca, \textit{An introduction
to Quantum Computing} (Oxford University Press, 2007), p. 146.

\end{document}